\documentclass[prd,preprint,tightenlines,floatfix,showpacs,preprintnumbers,nofootinbib,eqsecnum,superscriptaddress]{revtex4}
\usepackage[cp1250]{inputenc} 
\usepackage[T1]{fontenc} 
\usepackage{amsfonts} 
\usepackage{amsopn}   
\usepackage{amssymb}  
\usepackage{amsthm}   
\usepackage{graphicx}
\usepackage{mathtools}
\usepackage{enumerate}
\graphicspath{{../J_psi_ccbarg/old/}}

\newcommand{\bq}{\mbox{\boldmath $q$}}
\newcommand{\br}{\mbox{\boldmath $r$}}

\newcommand{\brho}{\mbox{\boldmath $\rho$}}

\newcommand{\bb}{\mbox{\boldmath $b$}}

\newcommand{\ket}[1]{| {#1} \rangle}
\newcommand{\bra}[1]{\langle {#1} |}
\newcommand{\ave}[1]{\langle {#1} \rangle}
\newcommand{\half}{{1\over 2}}

\def\lsim{\mathrel{\rlap{\lower4pt\hbox{\hskip1pt$\sim$}}
		\raise1pt\hbox{$<$}}}         
\def\gsim{\mathrel{\rlap{\lower4pt\hbox{\hskip1pt$\sim$}}
		\raise1pt\hbox{$>$}}}         

\begin{document}

\vfill
\title{Gluon shadowing in nuclei and the role of the $c \bar c g$ state in the coherent photoproduction of $J/\psi$ in nucleus-nucleus collisions}

\author{Agnieszka {\L}uszczak}
\email{agnieszka.luszczak@pk.edu.pl} 
\affiliation{
	T.~Kosciuszko Cracow University of Technology, PL-30-084 
	Cracow, Poland}

\author{Wolfgang Sch{\"a}fer}
\email{Wolfgang.Schafer@ifj.edu.pl} \affiliation{Institute of Nuclear Physics Polish Academy of Sciences, 
	ul. Radzikowskiego 152, PL-31-342 Cracow, Poland}

\date{\today}

\begin{abstract}
In this brief note, we confront our results
on diffractive photoproduction of $J/\psi$ mesons with the putative gluon shadowing ratio defined as $R_g= \sqrt{\sigma(\gamma A \to J/\psi A)/\sigma_{\rm IA.}}$, where $\sigma_{\rm IA}$ is the result in impulse approximation.
Building on our earlier description of the process in the color-dipole approach, where we took into account the rescattering of
$c \bar c$ states only, 
we demonstrate that the inclusion of $c \bar c g$-Fock states improves the description at small $x$ commonly associated with gluon shadowing.
\end{abstract}



\maketitle

\section{Introduction}

The recent measurements
\cite{ALICE:2012yye,ALICE:2021gpt,CMS:2016itn,Bursche:2018eni,Kryshen:2017jfz,CMS:2023snh} of exclusive production of $J/\psi$ mesons in ultraperipheral heavy-ion collisions at the LHC have attracted much attention.

At high energies, the diffractive photoproduction of the $J/\psi$ meson can be described in the color dipole picture, see for example numerous approaches \cite{Goncalves:2005yr, AyalaFilho:2008zr,Ducati:2013bya,Cisek:2012yt,Lappi:2013am,SampaiodosSantos:2014puz,Goncalves:2017wgg,Xie:2016ino,Kopp:2018xvu,Luszczak:2017dwf,Luszczak:2019vdc,
Henkels:2020kju}.
In the dipole approach the imaginary part of the diffractive production amplitude in forward direction takes the form
\begin{eqnarray}
\Im m \mathcal{A}(\gamma A \rightarrow V A ;W,\bq=0)  &=& 2 \int_0^1 dz \int d^2\br \, \Psi^*_V(z,\br) \Psi_\gamma(z,\br)  \, \int d^2 \bb \, \Gamma_A(x,\bb,\br).
\label{eq:amplitude}
\end{eqnarray}
Here $\Psi_\gamma, \Psi_V$ stand for the light-front wave functions (LFWFs) of the 
the virtual photon and $J/\psi$ meson respectively.
The crucial input is the scattering amplitude $\Gamma_A(x,\br,\bb)$ of a color dipole of size $r$ in
impact parameter space \cite{Nikolaev:1990ja,Mueller:1989st}.
Here the color dipole are treated as a concrete set of diffraction scattering eigenstates \cite{Good:1960ba,Miettinen:1978jb,Kolya_DSE,Nikolaev:1990ja}.
For heavy nuclei, the dipole amplitude can be written  in the exponential from familiar from Glauber theory \cite{Glauber}. 
\begin{eqnarray}
\Gamma_A(x,\bb,\br) = 1 -  S_A(x,\bb,\br)\, , \, \, {\rm{with} }\, \, S_A(x,\bb,\br) = \exp\Big[-\half \sigma(x,\br) T_A(\bb)\Big] \, .
\label{eq:Glauber}
\end{eqnarray}
Above, $T_A(\bb)$ is the optical thickness of the nucleus of mass number $A$ normalized
as $\int d^2\bb T_A(\bb) = A$.
The dipole amplitude depends on the $\gamma A$ per-nucleon cm-energy $W$ via $x = M_V^2/W^2$. Here $\sigma(x,\br)$ is the dipole cross section for the free nucleon target.
An important feature of diffractive production on the free nucleon is the ``scanning radius'' property of the overlap integral Eq.~\ref{eq:amplitude}, which 
entails that the process is dominated by a typical dipole size $r = r_s$, in this case $r_s \sim 1/m_c$, where $m_c$ is the charm quark mass. At small dipole sizes, we have
\begin{eqnarray}
    \sigma(x,\br) =  \frac{\pi^2}{N_c} \alpha_s(\mu^2) r^2 \, xg(x,\mu^2) ,\quad \mu^2 = \frac{A}{r^2}+\mu_0^2 \, ,
\end{eqnarray}
which leads to the diffractive cross section proportional to the square of the proton's gluon distribution \cite{Ryskin:1992ui},
\begin{eqnarray}
    \sigma(\gamma p \to J\psi p) \propto \sigma^2(x,r_s) \propto [x g(x,\mu^2)]^2 \, .  
\end{eqnarray}
Clearly this motivates an attempt to find information on the {\it{nuclear}} gluon distribution from diffractive $J/\psi$ production on nuclear targets.
The strongly absorbing nuclear target  however introduces another scale, the saturation scale (for a review see \cite{Gelis:2010nm})
\begin{eqnarray}
    Q_A^2(x,\bb) = {4 \pi^2 \over N_c} \alpha_s(Q_A^2) xg(x,Q_A^2) T_A(\bb),
\end{eqnarray}
in terms of which the dipole-nucleus S-matrix Eq.\ref{eq:Glauber} takes the form 
\cite{Mueller:1999wm}
\begin{eqnarray}
S_A(x,\br,\bb) = \exp\Big[ -{1 \over 8} Q_A^2(x,\bb) r^2\Big] \, ,
\end{eqnarray}
Therefore the scale
\begin{eqnarray}
r_A = {2 \sqrt{2} \over Q_A} \, ,
\end{eqnarray}
separates standard hard processes with $r_s \ll r_A$, for which the leading twist pQCD factorization works, from processes in the strongly absorptive saturation regime $r_s \gsim r_A$.

An estimate in Ref.\cite{Nikolaev:2003zf} gives $\ave{Q_A^2(\bb)} \sim 0.9 \, \rm{GeV}^2$ for $A^{1/3} = 6$ and $x \sim 0.01$.
For diffractive $J/\psi$ photoproduction on lead we are therefore in a regime $r_s/r_A \sim 0.2$, where we may regard the rescatterings included in the Glauber-Gribov form as a summation of higher-twist corrections, but not yet as a fully developed saturated regime.

In Ref.\cite{Luszczak:2019vdc} based on Glauber-Gribov rescattering of a color dipole were compared to the rapidity dependent cross sections for the exclusive production of $J/\psi$ in ultraperipheral collisions of lead ions.
A rather good agreement was obtained for various input dipole cross sections when compared to the results of LHCb, ALICE and CMS at forward rapidities $|y| \gsim 2$. 

\section{Beyond the $c \bar c$ rescattering}

The Glauber form of the dipole amplitude however applies only in range of not too small $x = x_A \sim 0.01$. 
Indeed, in the dipole picture of high-energy scattering, higher Fock states containing multiple additional gluons are responsible for the energy dependence of the dipole cross section. Multigluon states strongly ordered in light cone momentum fraction $z_n \ll z_{n-1} \ll \dots \ll z_1 \ll 1$ give rise to the BFKL evolution of the dipole cross section \cite{Mueller:1993rr,Nikolaev:1993ke,Nikolaev:1993th}.
The dipole model applies in a limit of large coherence length of the relevant Fock state. For example, in the limit of large photon energy $\omega$ in the rest frame of the nucleus,  the coherence length $l_c = 2 \omega /M_{c \bar c}^2$ for $c \bar c$ invariant mass  $M_{c \bar c} \sim M_{J/\psi}$ becomes much larger than the size of the nucleus $l_c \gg R_A$ \cite{Kopeliovich:1991pu,Nikolaev:1992si}. This justifies the application of the Glauber approach for color dipoles \cite{Mueller:1989st,Nikolaev:1990ja} for the $c \bar c$ component.
With increasing energy, the coherency condition $l_c \gg R_A$ will be satisfied not only by the $c \bar c$-state, but also by higher $c \bar c g, c\bar c gg, \dots$ states. In the language of Glauber--Gribov theory, these correspond to 
inelastic shadowing corrections induced by high--mass diffractive states. 
Below we restrict ourselves to $c \bar c g$ states to represent high mass diffractive contributions. This is motivated by a phenomenological success of color dipole approaches such as \cite{Bartels:1998ea,Golec-Biernat:2008mko} which can describe inclusive diffractive structure functions at HERA just including $q \bar q$ and $q \bar q g$ states. We therefore follow previous considerations for inclusive and diffractive DIS \cite{Nikolaev:2006mh}, where a similar approach had been adopted. We also omit possible contributions of $c \bar c q \bar q$-Fock states addressed in \cite{Nikolaev:2004yh}.

We very closely follow the approach outlined in \cite{Nikolaev:2006mh} adopted to the problem at hand in \cite{Luszczak:2021jtr}, the details must not be repeated here.

Assuming the Glauber form to hold at $x \sim x_A$, account for the $c \bar c g$ states leads to a dipole amplitude
\begin{eqnarray}
	\Gamma_A (x,\br,\bb) = \Gamma_A(x_A,\br,\bb) + \log\Big({x_A \over x}\Big) \Delta\Gamma_A(x_A,\br,\bb)  \, ,
	\label{eq:Gamma_full}
\end{eqnarray}
where 
\begin{eqnarray}
\Delta \Gamma_A (x_A,\br,\bb) &=&  \int d^2\brho_1 |\psi(\brho_1) - \psi(\brho_2)|^2 \Big\{ \Gamma_{A, c \bar c g} (x_A,\brho_1,\brho_2,\bb) - \Gamma_{A, c \bar c}(x_A,\br,\bb) \Big\}
\label{eq:Delta_Gamma}
	\end{eqnarray}
and 
\begin{eqnarray}
\Gamma_{A, c \bar c g} (x_A,\brho_1,\brho_2,\bb) &=&   \Gamma_A(x_A,\brho_1,\bb+{\brho_2 \over 2}) 
	+ \Gamma_A(x_A,\brho_2,\bb+{\brho_1 \over 2}) 
	\nonumber \\ 
	&& - \Gamma_A(x_A,\brho_1,\bb+{\brho_2 \over 2}) \Gamma_A(x_A,\brho_2,\bb+{\brho_1 \over 2})
\end{eqnarray}
The transverse separation of the gluon from quark/antiquark, encoded in the radial part of the $q \to q g$ light front WF:
\begin{eqnarray}
\psi(\brho) = {\sqrt{C_F \alpha_s} \over \pi} 
{\brho \over \rho^2} \, \mu_G \rho K_1(\mu_G \rho) \, 
\label{eq:gluon_WF}
\end{eqnarray}
In Eq.(\ref{eq:Delta_Gamma}) the integration extend over all dipole sizes, including the infrared domain of large dipoles, where perturbation theory does not apply. Here, we follow \cite{Nikolaev:1994vf,Nikolaev:2006mh} and introduce the parameter $\mu_G$ which ensures a the finite propagation radius $R_c = 1/ \mu_G \sim 0.2 \div 0.3 \, \mathrm{fm}$ of gluons. In this work, we employ $\mu_G = 0.7 \, \rm{GeV}$. We use the running coupling
\begin{eqnarray}
    \alpha_s(r) = \frac{4 \pi}{9} \frac{1}{\log \Big(\frac{C}{r^2 \Lambda^2} \Big)} ,
\end{eqnarray}
with $C = 1.5$, which we freeze for $r > 1/\mu_G$. It enters the WF of Eq.(\ref{eq:gluon_WF}) as $\alpha_s(\min \{r,\rho\})$.
In order to obtain the impulse approximation result, i.e. the production on a quasifree nucleon in the nucleus, we switch off the nonlinear piece in $\Gamma_{A,c\bar c g}$ and put 
\begin{eqnarray}
\Gamma_{\rm IA}(x_A,\br,\bb) = \half \sigma(x_A,\br) T_A(\bb) \, ,
\end{eqnarray}
We then follow \cite{Cisek:2012yt} and evaluate
\begin{eqnarray}
R_{\rm coh}(x) = {\displaystyle \int d^2\bb \, \Big| \bra{J/\psi} \Gamma_A(x,\br, \bb) \ket{\gamma}\Big|^2 \over
\displaystyle \int d^2\bb \, \Big| \bra{J/\psi} \Gamma_{\rm IA}(x,\br, \bb) \ket{\gamma}\Big|^2 } \, ,
\end{eqnarray}
and finally obtain the nuclear cross section as
\begin{eqnarray}
\sigma(\gamma A \to J/\psi A; W) = R_{\rm coh}(x) \, \sigma(\gamma p \to J/\psi p; W)  \, .
\label{eq:total_cross_sec}
\end{eqnarray}
The cross section $\sigma(\gamma p \to J/\psi p; W)$ is taken from our previous work \cite{Luszczak:2019vdc}.

\section{Comparison to experiment}
\label{sec:results}
In our numerical calculations we employed an easy to use dipole cross section designed for the use with heavy flavours from Ref.\cite{Golec-Biernat:2006koa}. The light front wave function overlap is taken from \cite{Kowalski:2006hc}, as in \cite{Luszczak:2019vdc} We refer the reader to these references for further details. 

\begin{figure}[!h]
	\begin{center}
	\includegraphics[width=.45\textwidth]{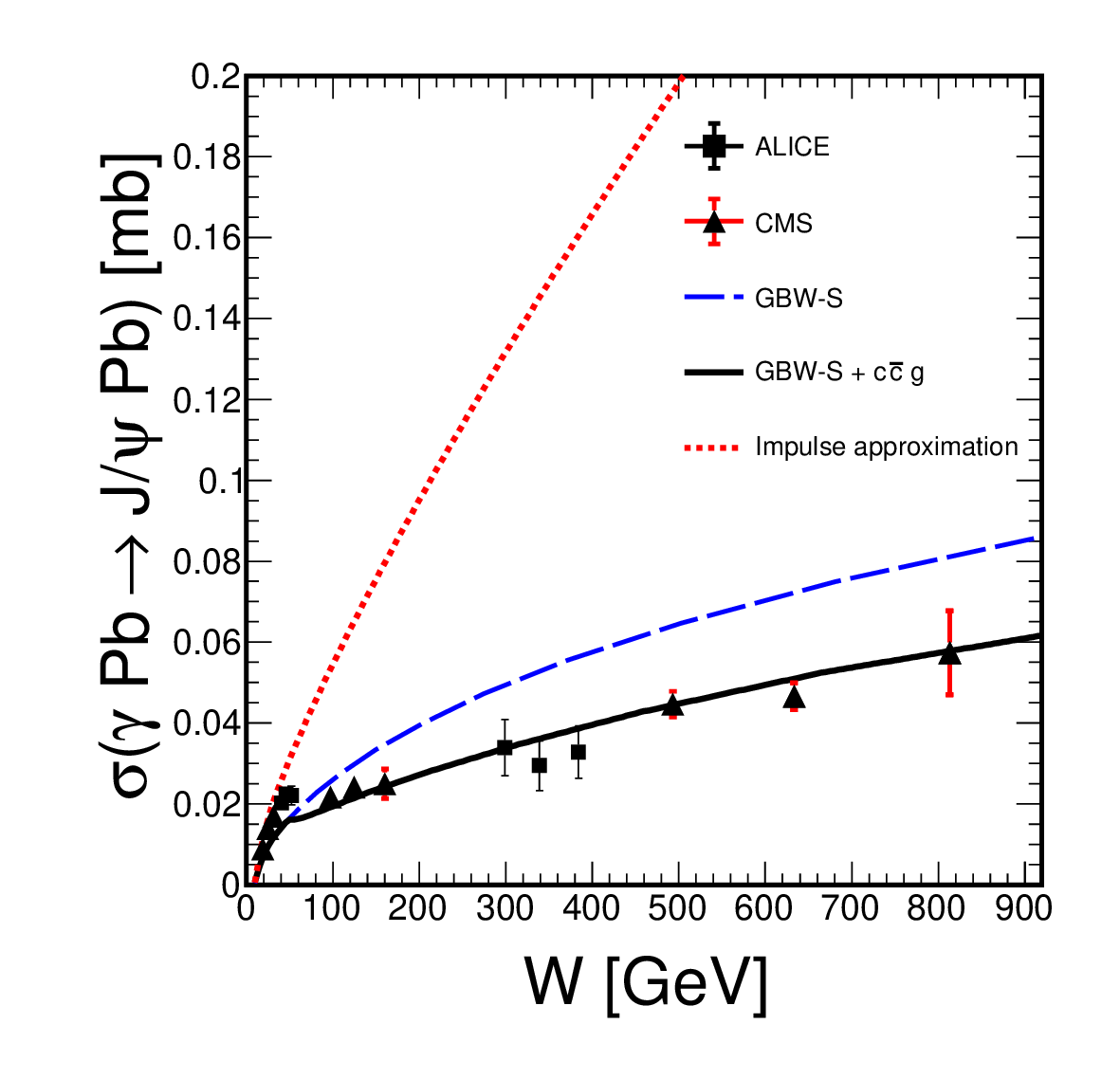}
		\caption{The total diffractive photoproduction
  cross section of $J/\psi$ on $^{208}\rm Pb$. The data points are taken from Refs.\cite{ALICE:2021gpt,CMS:2023snh,Contreras:2016pkc}  
		}
		\label{fig:sigma_A}
	\end{center}
\end{figure}

In Fig.\ref{fig:sigma_A} we show our results for the total diffractive photoproduction cross section of $J/\psi$ on lead as a function of $\gamma A$ per-nucleon cm-energy.
The data were obtained by the CMS \cite{CMS:2023snh} and ALICE \cite{ALICE:2021gpt} collaborations. We also show 
data points extracted by Contreras in Ref. \cite{Contreras:2016pkc} from
data obtained in ultraperipheral heavy-ion collisions.
We show separately, the result of the impulse approximation, the Glauber-Gribov rescattering of $c \bar c$ pairs, as well as the cross section including rescattering of the $c \bar c g$ state.
To be precise, the solid curve is obtained by adding
\begin{eqnarray}
\sigma(\gamma A \to J/\psi A; W) = (1 - f(W)) \sigma_{c \bar c} (\gamma A \to J/\psi A; W)  + f(W) \sigma_{c \bar c + c \bar c g} (\gamma A \to J/\psi A; W) \nonumber \\
\end{eqnarray}
with 
\begin{eqnarray}
    f(W) = \half \Big\{ 1 + \tanh\Big( \frac{W-W_0}{W_1} \Big) \Big\} \, ,
\end{eqnarray}
with $W_0 = 50 \, \rm{GeV}$, $W_1 = 5 \, \rm{GeV}$.
Clearly the impulse approximation fails dramatically, which illustrates the scale of nuclear effects. A large part of the nuclear suppression can be explained by Glauber-Gribov rescattering of the $c \bar c$ state alone.
We also observe that the calculations including the effect of the $c \bar cg$ state show an additional suppression of the nuclear cross section, as required by experimental data.

\begin{figure}[!h]
	\begin{center}
	\includegraphics[width=.45\textwidth]{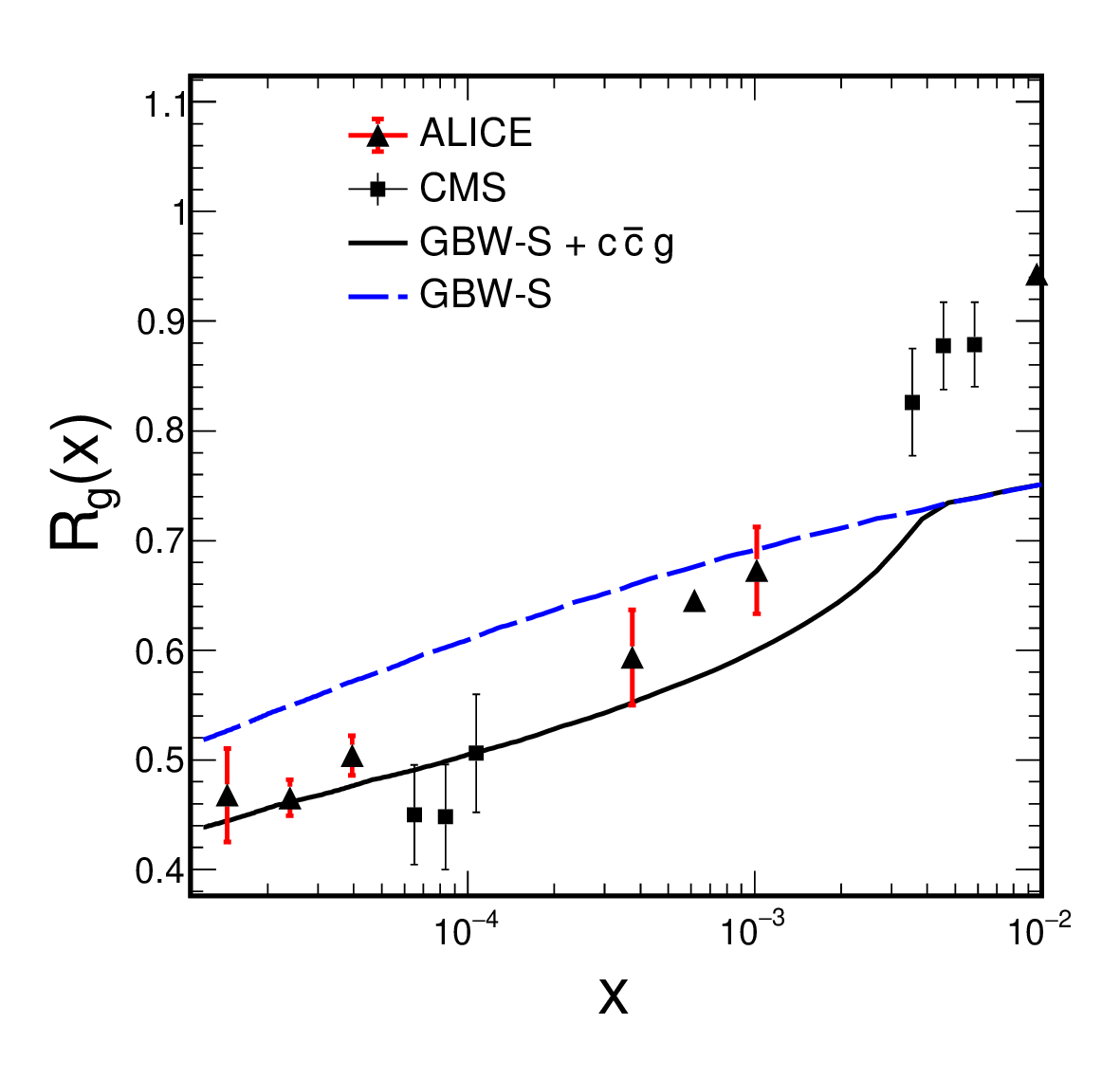}

		\caption{The ``gluon shadowing ratio'' $  R_g(x) = \sqrt{\frac{\sigma(\gamma A \to J/\psi A; W)}{\tilde \sigma_{\rm IA}(\gamma A \to J/\psi A; W)}} \, $ for $^{208}\rm Pb$. Notice that the reference cross section in the denominator is the parametrization taken from \cite{Guzey:2013xba}.
		}
		\label{fig:R_g}
	\end{center}
\end{figure}
Finally, in Fig.\ref{fig:R_g} we show the so-called gluon shadowing ratio $R_g(x)$.  The latter is defined as
\begin{eqnarray}
    R_g(x) = \sqrt{\frac{\sigma(\gamma A \to J/\psi A; W)}{\tilde \sigma_{\rm IA}(\gamma A \to J/\psi A; W)}} \, .
\end{eqnarray}
It is important to note, that here the impulse approximation cross section in the denominator, $\tilde \sigma_{\rm IA}$ is taken from Ref. \cite{Guzey:2013xba}, as this was also done in the analysis of the CMS collaboration \cite{CMS:2023snh}. 
This particular ratio is motivated by the leading-twist limit of the diffractive cross section, where it is expected that
\begin{eqnarray}
    R_g (x) \propto \frac{g_A(x,\mu^2)}{A g(x,\mu^2)} \, ,
\end{eqnarray}
and therefore $R_g(x)$ is supposed to quantify the suppression (shadowing) of the per-nucleon glue in the nucleus at small-$x$.
Here the scale $\mu^2 \sim M^2_{J/\psi}/4 \sim 2.4 \, \rm{GeV}^2$.

As can be seen from Fig.\ref{fig:R_g}, again including the $c \bar c g$ state contributes to an additional suppression improving the agreement with data. This is also in qualitative agreement with the results for the nuclear structure function $F_{2A}$ obtained in Ref. \cite{Nikolaev:2006mh}. 

At larger $x$, but still well within the range of applicability of the dipole approach we observe a large discrepancy with experimental data. Note that a similar discrepancy is also visible in the rapidity dependence of cross sections from ultraperipheral heavy-ion collisions (UPCs) shown in \cite{Luszczak:2019vdc,Luszczak:2021jtr}. There we found good agreement with LHCb data, which have not been analyzed in the form of Fig.\ref{fig:R_g}, but obtained a substantial underprediction of the ALICE forward rapidity data.

We note that this is the case for all dipole models used by us, and as can be seen from Ref.\cite{CMS:2023snh} the same is true for slightly different dipole model calculations by other groups. An analysis of exclusive $J/\psi$ production in UPCs based on the NLO collinear factorization approach \cite{Eskola:2022vaf} also finds an ambiguity of gluon distributions fitting either set of data.

From the point of view of the dipole approach some amount of overprediction of shadowing in the frozen dipole size approximation is indeed expected \cite{Zakharov:1998sv} in the range of $x \sim 0.01$.
We have accounted for the finite coherence length only in the most crude way, by including the nuclear form factor suppression \cite{Luszczak:2019vdc} depending on longitudinal momentum transfer.

\section{Conclusions}

In this letter we have addressed the diffractive photoproduction of $J/\psi$ mesons at the highest available energies.
Data extracted from UPCs are well described at high energies/small-$x$ after including additional shadowing that stems from the $c \bar c g$ Fock state.

Admittedly, our calculation relies on a model for the behaviour of the $c \to c g$ light front wave function in the nonperturbative region at large transverse distances. This is an inevitable element of any color dipole approach, though. We believe that our modeling of the essentially nonperturbative physics is well motivated by a phenomenological success of earlier works in the color dipole approach \cite{Nikolaev:1994vf,
Nikolaev:2006mh}.
Our restriction to the $c \bar c g$-system is to some extent justified by the fact, that diffractive structure functions of the proton measured at HERA are well described by the inclusion of $q \bar q$ and $q \bar q g$-states \cite{Bartels:1998ea,Golec-Biernat:2008mko}. 

It is interesting to ask, whether  our inelastic shadowing induced by $q \bar q g$ states can be directly related to a shadowing of the leading twist, DGLAP evolving gluon distribution of a nucleus.
While some early papers on the color dipole approach  \cite{Nikolaev:1993th,Nikolaev:1994kk,Zakharov:1998sv} indeed suggest so, such a conclusion would be ad odds with the treatment of parton saturation effects along the lines of \cite{Balitsky:1995ub,Kovchegov:1999yj}. A closer look at the triple Pomeron coupling \cite{Nikolaev:2006za,Bartels:2007dm} also casts some doubts on this interpretation.

A precise relation to gluon shadowing would require a removal of higher-twist corrections which are also present in our $c \bar cg$ contribution.
In any case, DGLAP evolution can only make sense at a hard scale $Q^2$ much larger than the saturation scale. In the case of gluon shadowing, the relevant saturation scale is the one for an gluon-gluon dipole, which is enhanced by a factor $C_A/C_F= 9/4$ over the one of the triplet-antitriplet dipole. Roughly, one expects $\ave{Q^2_{8A}(\bb)} \sim 2 \, \rm{GeV}^2$, in a similar ballpark as the hard scale for $J/\psi$ production $Q^2 \sim M^2_{J/\psi}/4 \sim 2.4 \, \rm{GeV}^2$. 
\bibliography{ccbarg}

\end{document}